\documentclass[a4paper,11pt]{article}

\usepackage[dvips]{graphicx}
\usepackage{verbatim}
\usepackage{latexsym,amsfonts,amssymb}
\usepackage[normalem]{ulem}
\usepackage{color}

\usepackage{colortbl}
\usepackage{url} 
\usepackage{caption}
\usepackage{verbatim}
\usepackage{fancyvrb}
\usepackage{subfigure}
\usepackage{calrsfs}
\usepackage{amsmath}

\begin{document}

\begin{center}
\ \\\vspace{7cm}
\noindent\textbf{\LARGE Smoothing and differentiation of data by Tikhonov and fractional derivative tools, {{applied to surface-enhanced Raman  scattering (SERS) spectra of crystal violet dye 
}}}\\
\vspace{2cm}
{\small\noindent{\bf Nelson H. T. Lemes$^{1,*}$,Tain\'ah M. R.
Santos$^{2}$, Camila A. Tavares$^{2}$, Luciano S. Virtuoso$^{3}$, 
Kelly {A. S. Souza}$^{3}$}, {\bf Teodorico C. Ramalho}$^{2}$} \\
\vspace{2cm}
\end{center}

\noindent
\begin{center}
{\small $^1$ 
Laboratory of Mathematical Chemistry, Institute of Chemistry,\\
Federal University of Alfenas,
Alfenas/MG, 37130-001, Brazil\\
$^2$ 
Laboratory of Molecular Modeling, Department of Chemistry, \\
Federal University of Lavras,
Lavras/MG, 37200-900, Brazil\\
$^3$  
{Laboratory of Colloid Chemistry}, Institute of Chemistry, \\
Federal University of Alfenas,
Alfenas/MG, 37130-001, Brazil
}\\
\end{center}
\vspace{\stretch{1}}
\noindent{$^*$nelson.lemes@unifal-mg.edu.br}

\newpage
\subsubsection*{Abstract}

All signals obtained as instrumental response of analytical apparatus are affected by noise, as in Raman spectroscopy. Whereas Raman scattering is an inherently weak process, the noise background can lead to misinterpretations. Although surface amplification of the Raman signal using metallic nanoparticles has been a strategy employed to partially solve the signal-to-noise problem, the pre-processing of Raman spectral data through the use of mathematical filters has become an integral part of Raman spectroscopy analysis. In this paper, a Tikhonov modified method to remove random noise in experimental data is presented. In order to refine and improve the Tikhonov method as filter, the proposed method includes Euclidean norm of the fractional-order derivative of the solution as an additional criterion in Tikhonov function. In the strategy used here, the solution depends on the regularization parameter, $\lambda$, and on the fractional derivative order, $\alpha$. As will be demonstrated, with the algorithm presented here, it is possible to obtain a  noise free spectrum without affecting the fidelity of the molecular signal. In this alternative, the fractional derivative works as a fine control parameter for the usual Tikhonov method. The proposed method was applied to simulated data and to surface-enhanced Raman scattering (SERS) spectra of crystal violet dye in Ag nanoparticles colloidal dispersion.\\

\noindent
{\bf Keywords:} Fractional derivative; Tikhonov filter; Raman  spectroscopy; SERS effect; silver nanoparticles.

\clearpage

\section{Introduction}

Often in chemical research it is common to obtain in single analysis a large sequence of measurements 
that are related by some variable, such as wave number. For example, Raman and IR spectroscopy provide detailed chemical information from radiated energy as function of the wave number. These techniques are routinely used in various application areas including pharmaceutical, polymers, forensic, environmental, food sciences, biological diagnostic, etc. \cite{Barnett2014,Gahlaut2020}. The experimental IR or Raman spectra usually contain noise that can lead to misinterpretation. In this case, analytical signal processing techniques are used to remove the noise from the experimental data \cite{Rinnan2009,Ahlinder2016,Barton2018,Bai2020a}. The goal of these mathematical operations is to remove noise without losing much details in experimental data, which could be correlated to some physical or chemical property of interest. 

These techniques, widely applied in analytical chemistry, are denominated mathematical filters. The most common is the polynomial least-squares smoothing filter, more commonly known as Savitzky-Golay filter \cite{Merkle1964,Schafer2011}. Another very common technique is based on Fourier transform and inverse Fourier transform, called Fast-Fourier transform filter \cite{Zhao2014}, in which the separation of the frequency components of the Raman spectrum from those noises is carried out.
 
A lesser known method in chemistry analysis is Tikhonov algorithm. Although widely used in image processing and inverse problems, its use in chemometrics is not often found in literature, some examples are \cite{Aydin2008,Liu2013,Liu2015,Gutta2018,Skogholt2018}. As it will be shown, the Tikhonov method (TM) is easy to implement and efficient. In Tikhonov method, the regularization
parameter controls the weight given to each information in the objective function. Therefore,  choosing the regularization parameter controls very well how much signal is removed. As previously highlighted, the control of how much or how little noise is removed from the data signal, without affecting the main signal, is very important to avoid misinterpretation. The Tikhonov method has
been a very powerful tool in filtering out the influence of noise and fluorescence background in Raman spectrum \cite{Liu2013}. In reference \cite{Liu2013}, the authors proposed a model that combines baseline correction and denoising, performing both operations simultaneously. However, as we shall see, more efficient methods are needed in order to  minimize oscillations in the solution and 
allow greater control of the filter.

In a attempt to refine and improve the Tikhonov method \cite{1,Tikhonov1,Tikhonov2} as filter, a new denosing method based in Tikhonov method is presented in this paper. The proposed method includes Euclidean norm of the fractional-order derivative of the solution as an additional criterion in the objective function. In this paper, Gr\"unwald-Letnikov fractional derivative will be used to generalize the derivative concept from integer order to fractional order \cite{dfracgl1,dfracgl2,appdfrac1,appdfrac2}. Recently, many papers in literature have reported the use of fractional-order derivative as a better alternative than to integer-order derivative \cite{appdfrac3,Nelson1,Nelson2}. A similar improvement of the Tikhonov regularization method was proposed by us and tested in the inverse black body radiation problem. This work shows that the proposed modification is more flexible than the original method, with better results using fractional derivative order instead of integer order \cite{Nelson1}. However, the improvement in Tikhonov method as filter technique,  proposed here,  has not yet been explored in the literature. In this paper, the Fractional Tikhonov method will be presented and used in the pre-processing of experimentally Raman spectra obtained for crystal violet dye solutions in contact with a dispersion of silver nanoparticles. An evaluation of the quality of the pre-treatment will be performed by comparing the result obtained by Tikhonov method with that obtained by the modified Tikhonov method with fractional derivative operator in objective function. 

Of the many definitions for fractional order derivative, the Gr\"unwald-Letnikov (GL) definition is the most often used \cite{dfracgl1}. The Gr\"unwald-Letnikov definition is simply a generalization of the ordinary discretization formula for integer order derivatives and most straightforward from the point of view of numerical implementation, being numerically equivalent to a Riemann-Liouville derivative. For this purpose, the matrix form of Gr\"unwald-Letnikov will be used.
The fractional calculus provides an excellent instrument for the description of memory properties of various processes \cite{dfracgl1}, which is the motivation for this study. The use of the GL derivative in Tikhonov function was first presented in the reference \cite{Nelson1} in the context of the study of inverse problem. Now  Fractional Tikhonov Method (FTM) will be used as filter to remove noise from Raman spectrum.

Particularly, with the advent of the development of Surface-Enhanced Raman Scattering (SERS), which took place in the 1970s, research in the areas of Physics and Chemistry of interfaces has advanced significantly with the emergence of several potential applications of this technique, ranging from the development of new sensors to biomedical applications \cite{Marzan2020}.

In addition to what has been said before, the Raman spectroscopy will be chosen to test the proposed method because Raman scattering is an inherently weak process, which often results in low signal, meaning that Raman spectroscopy is vulnerable to noise \cite{Wentzell2000}. These noises arise from several sources, which  can be reduced by experimental control procedure. However, these measures can be inefficient or can be costly in terms of time and equipment and thus are impractical in certain applications. Therefore, an efficient denoising algorithm for mathematical processing of spectra would be advantageous.

In the strategy proposed here, the solution will depend on the regularization parameter, $\lambda$, and on the fractional derivative order, $\alpha$. For $\alpha$ assuming real value between 0 and 2, the solution obtained is more adequate than the ones from the usual Tikhonov regularization method.
Using this technique, it will be possible to obtain a smooth noise free spectrum without affecting the fidelity of sharp spectral features in this process. In this work, the background line will be considered as part of useful signal, because random noise reduction is the major focus of this proposed method. In the future, modified procedures based on this proposal may be implemented, in which baseline removal and noise reduction are carried out simultaneously.

In section 2, we will begin by introducing our generalized model and the necessary foundations of fractional calculus will be presented. Details of the experimental procedure to obtain the SERS spectra are going to be presented in section 3. The prototype problem for the examination of the method proposed is going to be presented in section 4 together with results obtained and discussion of the efficiency of the strategy proposed here. Then, in section 5, the conclusions will be summarized.

\section{Methodology}

\subsection{Generalized Tikhonov functional}
\label{dfratik}

When experimental ${f}(\nu)$ function contains noises which often oscillate with large amplitude and frequency, the filter algorithms are needed to provide a clear information about the features of the observed system. Tikhonov method can be used as a strategy to solve this problem, with success in many cases \cite{Aydin2008,Liu2013,Liu2015,Gutta2018,Skogholt2018}. The Tikhonov method consists of finding the solution which minimizes the function $\Phi_\lambda({\bf f})$, where $\Phi_\lambda({\bf f})$ is given by \cite{Tikhonov1}
\begin{equation}
\Phi_\lambda({\bf f})=||{\bf f}-{\bf \hat f}||^2
+\lambda^2\left(a_0||{\bf f}||^2+a_1||{\bf f}^\prime||^2+a_2||{\bf f}^{\prime\prime}||^2\right)
\label{defderi}
\end{equation}
Here, the prime symbol is the first-order derivative of ${\bf f}$ and the double prime symbol is used to represent the second order derivative of ${\bf f}$. The ${\bf \hat f}$ function is the experimental data point \cite{1,Tikhonov1,Tikhonov2}.

Along this path, the strategy is to recover input signal without noise. Unlike the usual Tikhonov method in which the goal is to find ${\bf g}$, such that ${\bf Kg}={\bf f}$, in this case the output ${\bf f}$ is a new  ${\bf \hat f}$. Therefore, the solution ${\bf f}$ is experimental data ${\bf \hat f}$ without noise.

Parameters $a_0$, $a_1$ and $a_2$ are keys, which are used to lock ($a_i=1$) and unlock ($a_i=0$) the additional restriction on square difference norm. The regularization parameter $\lambda$ represents the weight given to the additional  restriction, therefore the solution provided  depends on the chosen regularization parameter. 

The most simple way to obtain the differential operator is an obvious method based on the definition of $df(\nu)/d\nu$ and $d^2f(\nu)/d\nu^2$, in this case
\begin{equation}
\left.\frac{df}{d\nu}\right|_{\nu_i}=\lim_{\Delta \nu\rightarrow 0}\frac{\Delta f(\nu)}{\Delta \nu}\approx \frac{f(\nu_{i}+h)-f(\nu_i)}{h}
\end{equation}
and
\begin{equation}
\left.\frac{d^2f}{d\nu^2}\right|_{\nu_i}=\lim_{\Delta \nu\rightarrow 0}\frac{\Delta f^\prime(\nu)}{\Delta \nu}\approx \frac{f^\prime(\nu_{i}+h)-f^\prime(\nu_i)}{h}\approx \frac{f(\nu_i+2h)-2f(\nu_i+h)+f(\nu_i)}{h^2}
\end{equation}
in which $\Delta \nu=h$, then Equation (\ref{defderi}) can be written as
\begin{equation}
\Phi_\lambda({\bf  f})=
\sum_i^n ({f}_i-\hat{f}_i)^2
+\lambda^2\left[a_0\sum_i^n f_i^2
+a_1\sum_i^{n-1}\frac{(f_{i+1}-f_i)^2}{h^2}
+a_2\sum_i^{n-2}\frac{(f_{i+2}-2f_{i+1}+f_i)^2}{h^4}\right]
\end{equation}
Following the steps given in \cite{riele}, but now deriving with respect to ${f}_i$, we obtain
\begin{equation}
\begin{array}{c}
\frac{\partial \Phi_\lambda}{\partial f_i}=
2(f_i-\hat{f}_i)
+\lambda^2\left[a_02f_i\right.
-\frac{a_12}{h^2}(f_{i+1}-f_i)
+\frac{a_12}{h^2}(f_i-f_{i-1})\\
\left.+\frac{a_22}{h^4}(f_{i+2}-2f_{i+1}+f_i)
-\frac{a_24}{h^4}(f_{i+1}-2f_{i}+f_{i-1})
+\frac{a_22}{h^4}(f_{i}-2f_{i-1}+f_{i-2})
\right]=0
\end{array}
\end{equation}
after a rearrangement, we obtain
\begin{equation}
\frac{1}{2}\frac{\partial \Phi_\lambda}{\partial f_i}=
(f_i-\hat{f}_i)
+\lambda^2\left\{a_0f_i-\frac{a_1}{h^2}(f_{i+1}-2f_i+f_{i-1})
+\frac{a_2}{h^4}(f_{i+2}-4f_{i+1}+6f_i-4f_{i-1}+f_{i-2})\right\}
=0
\end{equation}
and, finally, the equation 
\begin{equation}
{\bf \hat f}=
\left[\lambda^2\left\{a_0{\bf H}_0+\frac{a_1}{h^2}{\bf H}_1+\frac{a_2}{h^4}{\bf H}_2\right\}+{\bf H}_0\right]{\bf f}
\label{eqtik}
\end{equation}
if ${\bf a}_0=0$ so
\begin{equation}
{\bf f}_\lambda^*=\left[\lambda^2\left\{\frac{a_1}{h^2}{\bf H}_1+\frac
{a_2}{h^4}{\bf H}_2\right\}+{\bf H}_0\right]^{-1}{\bf \hat f}
\label{tikusual}
\end{equation}
in which ${\bf f}^*_\lambda={\bf f}$, ${\bf H}_{0}={\bf I},$
\begin{equation}
{\bf H}_{1}=\left[\begin{array}{ccccc}
1 & -1 & & &  \\
-1 & 2 & -1&  &\\
&\ddots &\ddots &\ddots  & \\
& & -1 & 2 & -1\\
&  &   & -1&1\\
\end{array}\right]
\end{equation}
\begin{center}
and
\end{center}
\begin{equation}
{\bf H}_{2}=\left[\begin{array}{ccccccc}
1 & -2 &1 & & & &\\
-2 & 5 & -4& 1& &&\\
1&-4&6&-4&1& &\\
&\ddots &\ddots &\ddots &\ddots &\ddots &\\
&&1 &-4 &6 &-4 &1\\
&&&1& -4 &5 &-2\\
&& & &1 & -2 & 1\\
\end{array}\right].
\end{equation}
If derivative operator of first-order is given by
\begin{equation}
{\bf D}=\left[\begin{array}{ccccccc}
-1 & 1 & 0 & 0 & ...&0 &0 \\
0 & -1 & 1 & 0 & ... & 0&0 \\
 & & &\vdots & & & \\
0 & 0 & ... & 0 & -1 & 1 & 0 \\
0 &0 & ... & 0 & 0 & -1 & 1
\end{array}\right]
\label{oped}
\end{equation}
in Equation (\ref{eqtik}), ${\bf H}_{0}$ is the identity matrix ${\bf I}$, ${\bf H}_{1}={\bf D}^T{\bf D}={\bf D}^2$ 
and ${\bf H}_{2}=({\bf D}^T{\bf D})^T({\bf D}^T{\bf D})={\bf D}^4$ \cite{riele,apptik1}.
If the vector {\bf f} is defined by
${\bf f}^*_\lambda=[f^*_\lambda(\nu_1), f^*_\lambda(\nu_2), f^*_\lambda(\nu_3), ...]^T$  
in which $\nu_i=\nu_{i-1}+h$ then
\begin{equation}
{\bf f}_\lambda^*=\left[\lambda^2\left\{\frac{a_1}{h^2}{\bf D}^2+\frac
{a_2}{h^4}{\bf D}^4\right\}+{\bf I}\right]^{-1}{\bf \hat f}
\label{eqtik2}
\end{equation}
Notice that the Tikhonov method makes no assumptions about the function shape $f(\nu)$. 

When $\lambda$ is equal to zero, the solution found minimizes the difference between ${\bf \hat f}$ and ${\bf f}^*_0$. In this case, the solution that minimizes the cost function (Equation (1)) is ${\bf f}^*_0={\bf \hat f}$. Therefore, the noise is not removed. On the other hand, if $\lambda$ is large, we find a solution with little oscillations, i.e. with little values of the norm of the ${\bf f^\prime}$, but the norm of the difference between ${\bf \hat f}$ and ${\bf f}^*_\lambda$ is quite large. Thus, there is one $\lambda$ that establishes the good balance between staying close of the experimental data and free of large oscillations. In this paper, the modified L-curve \cite{Lcurve} was used to get the optimal regularization parameter. The modified L-curve is a plot of the norm of first-order derivative of the regularized solution $-$ $||{\bf f}^{*(1)}_\lambda||$ $-$ versus the norm of the corresponding difference  between regularized solution and experimental signal $-$ $||{\bf f}^*_\lambda-{\bf \hat f}||$. The L-curve is a well-known heuristic method in which optimal regularization parameter is obtained by selecting the corner in the curve shape (maximum curvature) \cite{Lcurve}.

The solution presented in Equation (\ref{eqtik2}) is equal to solution of the system of linear equations below \cite{apptik1} 
\begin{equation}
\left[
\begin{array}{c}
{\bf I}\\
{\bf L}_1\\
{\bf L}_2
\end{array}\right]{\bf f}_\lambda^*=
\left[
\begin{array}{c}
{\bf \hat f}\\
{\bf 0}
\end{array}\right]
\end{equation}
where ${\bf L}_1=\lambda\frac{\sqrt{a_1}}{h}{\bf D}$ and ${\bf L}_2=\lambda\frac{\sqrt{a_2}}{h^2}{\bf D}^2$. This strategy can be improved by using non-integer derivative operator, then 
\begin{equation}
\left[
\begin{array}{c}
{\bf I}\\
{\bf L}_\alpha
\end{array}\right]{\bf f}_{\lambda,\alpha}^*=
\left[
\begin{array}{c}
{\bf \hat f}\\
{\bf 0}
\end{array}\right]
\end{equation}
where ${\bf L}_\alpha=\lambda\frac{1}{h^{\alpha}}{\bf D}^{\alpha}$
{by procedures similar to those for Equation (\ref{eqtik2}) we obtain }
\begin{equation}
{\bf f}_{\lambda,\alpha}^*=\left[\frac{\lambda^2 }{h^{2\alpha}}{\bf D}^{2\alpha}+{\bf I}\right]^{-1}{\bf \hat f}
\label{eqprincipal}
\end{equation}
where ${\bf D}^{2\alpha}$ is the Gr\"unwald-Letnikov operator for the fractional derivative and $2\alpha$ is the derivative order \cite{dfracgl1}. Now regularized solution depends on the regularization parameter, $\lambda$, and fractional derivative order, $2\alpha$. The Gr\"unwald-Letnikov fractional derivative operator has not been used before in the present context. The question about how $\alpha$ has an effect on solution ${\bf f}_{\lambda,\alpha}^*$, at $\lambda$ fixed, will be discussed in section 4.

When $\alpha=1$, the solution from Equation (\ref{eqprincipal}) is equal to the case in which $a_1=1$ and $a_2=0$ in Equation (\ref{eqtik2}). The solution of Equation (\ref{eqtik2}), when $a_1=0$ and $a_2=1$, provides the same result that $\alpha=2$ in Equation (\ref{eqprincipal}). Finally, Equation (\ref{eqtik2}) and Equation (\ref{eqprincipal}) lead to the same result when  $a_1=0$ and $a_2=0$, with $\alpha=0$ in derivative order. Out of these special cases, there are many solutions which are not contained in solution space of Equation (\ref{eqtik2}), but are contained in solution space of Equation (\ref{eqprincipal}) in which $\alpha\in \mathbb{R}^+$ and $0\leq\alpha \leq 2$. Details of the Gr\"unwald-Letnikov derivative will be presented in subsection 2.3.

\subsection{Derivative filter}

The numerical differentiation of  numerical data is a common practice in analytical chemistry. This procedure can be used, for example, to locate the position of a maximum peak and highlight poorly defined features in a signal data \cite{Wentzell2000}. Whereas these measurements yield intrinsic errors, if we use finite difference method as an approximation to calculate derivative value, 
then arbitrarily small noise in the experimental data can lead to arbitrarily large perturbations in the result. In other words, the function $\hat{{\bf f}}^\prime$ does not depend continuously on $\hat{{\bf f}}$, which is a well-known criterion of the ill-posed problem \cite{apptik1}. Therefore, the  calculation of the $n$th order derivative requires some form of regularization to extract meaningful information. 

Tikhonov regularization was first applied to the numerical differentiation problem in {\cite{[5]}}, in which the regularized derivative function of the $n$th order, ${\bf f}_\lambda^{*(n)}$, is given by
\begin{equation}
{\bf f}_\lambda^{*(n)}=\left[\lambda^2\left\{\frac{a_1}{h^2}{\bf D}^2+\frac
{a_2}{h^4}{\bf D}^4\right\}+{\bf I}\right]^{-1}{\bf D}^{n}{\bf \hat f}
\end{equation}
with $n=1,2,3,...$ Similar to what was done in subsection \ref{dfratik}, Tikhonov method was modified with fractional derivative operator in Tikhonov objective function, leading to the equation 
\begin{equation}
\label{filtrod}
{\bf f}^{*(n)}_{\lambda,\alpha}=\left[\frac{\lambda^2}{h^{2\alpha}}{\bf D}^{2\alpha}+{\bf I}\right]^{-1}{\bf D}^{n}{\bf \hat f}
\end{equation}
where ${\bf D}^{2\alpha}$ is  Gr\"unwald-Letnikov fractional derivative of order $2\alpha$  \cite{dfracgl1}. When the Equation (\ref{filtrod}) is used to find ${{\bf f}}^{(n)}$, we have a new parameter, that is $\alpha$. In this case, with fixed regularized parameter $\lambda$, we can use $\alpha$ to control the amplitude of the oscillations in result.

In general, Raman spectroscopy signal includes a strong fluorescence background, which can mask the weak Raman signal. Therefore the fluorescence signal in Raman spectroscopy directly affects the accuracy and sensitivity of Raman measurements. The derivative method, shown previously, has been widely described as method for fluorescence suppression, in which taking the derivative of a measured Raman spectrum will eliminate the fluorescence background. As mentioned above, high-frequency noises are often amplified by this method. For this reason, the proposed method  can be useful in new strategies to remove baseline background \cite{novoluc}.

\subsection{Fractional calculus background}

The Gr\"unwald-Letnikov $\alpha$th-order fractional derivative of a function $f(\nu)\in \mathbb{R}^n$ with respect to $\nu\in(0,\nu_{max}]$ is given by  a generalization of the formula for the  $n$th
derivative of $f(\nu)$, with $n$ = 1, 2, 3 ... \cite{dfracgl1},
\begin{equation}
[^{GL}D_0^{(\alpha)} f](\nu)=
\lim_{h\rightarrow 0^+} \frac{1}{h^\alpha}
\sum_{k=0}^{N} (-1)^k \left(
\begin{array}{c}
\alpha\\
k
\end{array}
\right) f(\nu-kh)
\label{eqdglx}
\end{equation}
where $n$ was replaced by fractional order $\alpha$ ($\alpha \in \mathbb{R}^+$) and the binomial coefficient was written as, 
\begin{equation}
\left(
\begin{array}{c}
\alpha\\
k
\end{array}
\right)=C_{\alpha,k}=
\frac{\Gamma(\alpha+1)}{\Gamma(k+1)\Gamma(\alpha-k+1)}
\label{bincoeffx}
\end{equation}
In Equation (\ref{bincoeffx}), the symbol $\Gamma(.)$ is the Euler's gamma function, an interpolation function for the factorial $n!$ when the argument $n+1$ is a non-integer number. From Equation (\ref{bincoeffx}), binomial coefficients can be calculated also to non-integer order $\alpha$. The total terms in the sum on the Equation (\ref{eqdglx}), $N=\lceil \nu/h\rceil$, is the smallest integer such that $N > \nu/h$. The sum shown in Equation (\ref{eqdglx}) converges absolutely and uniformly for each $\alpha > 0$ and for every bounded function $f(\nu)$.

The Gr\"unwald-Letnikov fractional derivative of $f(\nu)$, $[^{GL}D_0^{(\alpha)} f](\nu)$, 
shown in Equation (\ref{eqdglx}), has also a matrix form representation, such as ${\bf d}={\bf D}^{(\alpha)}{\bf f}$, with \cite{dfracgl1}
\begin{equation}
{\bf D}^{(\alpha)}=\frac{1}{h^{\alpha}}
\left[\begin{array}{cccccc}
C_{\alpha,0} & 0 & 0 & 0 &   ...&  0\\
-C_{\alpha,1} & C_{\alpha,0} & 0 & 0 &   ...&  0\\
C_{\alpha,2} & -C_{\alpha,1} & C_{\alpha,0} & 0 &  ...&  0\\
... & ... & ... & ...  & ... &  ...\\
(-1)^NC_{\alpha,N} & (-1)^{N-1}C_{\alpha,N-1} & (-1)^{N-2}C_{\alpha,N-2} & (-1)^{N-3}C_{\alpha,N-3}&
 ... & C_{\alpha,0}\\
\end{array}\right]
\end{equation}
${\bf f}=[f(0) f(h) f(2h) ... f(Nh)]^T$ and ${\bf d}=[[^{GL}D_0^{(\alpha)} f](0) [^{GL}D_0^{(\alpha)} f](h) ... [^{GL}D_0^{(\alpha)} f](Nh)]$. A proof of this result can be found in reference \cite{dfracgl1}. Here recursive formula was used to calculate binomial coefficients \cite{dfracgl1}
\begin{equation}
w_k^{(\alpha)}=\left(1-\frac{\alpha+1}{k}\right)w_{k-1}^{(\alpha)}
\end{equation}
where $w_k^{(\alpha)}=(-1)^k C_{\alpha,k}$. The recurrence relation provides a stable means for computing $C_{\alpha,k}$, when $k$ is large.

Some properties of Gr\"unwald-Letnikov (GL) fractional derivative are \cite{dfracgl1}: a) the GL fractional derivative of order zero returns the original function; b) the GL fractional derivative of integer order $n$ gives the same result as the usual differentiation of order $n$ and c) the GL fractional derivative is a non-local operator, as it is evident from sum in Equation (\ref{eqdglx}). 
The non-local property of the fractional operators is presented in the solution provided by Equation (\ref{eqprincipal}), which does not exist in the solution found using Equation (12). This fact provided the motivation for a change in Tikhonov method based on derivative with non-integer order.

\section{Experimental data}

This section describes how the experimental data was obtained to verify the performance of the proposed method.

\subsection{Materials}
\indent

Silver nitrate $(AgNO_{3})$, tribasic sodium citrate  $(Na_{3}C_{6}H_{5}O_{7}.2H_{2}O)$  and  sodium  lauryl sulfate $(NaC_{12}H_{25}SO_{4})$ acquired by Vetec Qu\'{\i}mica Fina. Sodium hydroxide $(99\%, NaOH)$ purchased from Sigma-Aldrich. Sodium borohydride $(NaBH_{4})$ from Nuclear. Acetone $(C_{3}H_{6}O)$ from Alphatec. Violet crystal P.A $(C_{25}H_{30}N_{3}Cl)$ acquired by Din\^amica Qu\'{\i}mica Contempor\^anea Ltda. Milli-Q water obtained by a Millipore purifier.

\subsection{Synthesis of silver nanoparticles}
\indent

The synthesis of $Ag$ nanoparticles ($AgNPs$) was performed according to the protocol described by L. M. Moreira, et al \cite{Agsint}. The molar ratio of $Ag^{+}$:citrate:borohydride was 1:1:3, respectively. A $1.0 \times 10^{-3}$ mol L$^{-1}$ $AgNO_{3}$ stock solution was prepared. The stock solution was diluted to a final concentration of $5.0 \times 10^{-4}$ mol L$^{-1}$ and added to a three-way flask coupled to a reflux condenser. The solution was brought to boiling and stirring and $0.2$ mL of $4.8 \times 10^{-4}$ mol L$^{-1}$ NaOH solution was added. After adding the NaOH solution, stirring and boiling were continued for $5$ minutes. Subsequently, $5$ mL of sodium citrate solution $5.0 \times 10^{-4}$ mol L$^{-1}$ was quickly added. After addition of citrate, the suspension turned yellowish. Then, $5$ mL of a $1.5 \times 10^{-3}$ mol L$^{-1}$ $NaBH_{4}$ solution was slowly dripped into the main solution. After that, the system was refluxed and stirred for another 15 minutes and then cooled to room temperature. The final suspension obtained had a caramel color.

\subsection{SERS Measurements}
\indent

To obtain the Raman spectra, aqueous colloidal dispersions were prepared containing $AgNPs$ ($9.97 \times 10^{-4}$ mol L$^{-1}$); the surfactant sodium lauryl sulfate (SDS), at a concentration of $4.0 \times 10^{-3}$ mol L$^{-1}$ and the crystal violet dye ($CV$), at concentrations of $1.0 \times 10^{-4}$ mol L$^{-1}$ to $1.0 \times 10^{-7}$ mol L$^{-1}$. The Raman spectra of all samples were obtained using a Raman spectrometer $i-HR550$ with a CCD detector, manufactured by Horiba (USA). The excitation wavelength used was $532$ nm and a $10\times $ objective. The samples were analyzed with a retention time of $60 s$, a multiplier of $3$ with a maximum laser power of $1$ mW.

\section{Results and discussions}

\subsection{Simulated data}
\label{simulate}

In this subsection, simulated data with known background and peaks are used to evaluate the proposed method (FTM) in comparison to usual method (TM). Because random noise reduction is the major focus of this method, the background line will be considered as part of useful signal. In the future, modified procedures based  on this proposal  may be implemented,  in which baseline removal and noise reduction is performed simultaneously.

The simulated data is intended to mimic the experimental data that contain analytical signal, fluorescence background and random noise. In this subsection, Raman spectra with different baselines and
noise were simulated and used to evaluate our method with regard to two different aspects: (i) norm of the first-order derivative of the recovered signal, $||{\bf f}^\prime||$; and (ii) norm of the difference between the analytical signal and recovered signal, $||{\hat{\bf f}}-{\bf f}||$.

Therefore, the simulated Raman spectrum can be modeled as 
\begin{equation}
\hat{\bf f}={\bf f_0}+{\bf b}+{\bf n},
\end{equation} 
in which ${\bf b}$ is the baseline of the Raman spectrum ${\bf f_0}$ and ${\bf n}$  is the experimental noise. The pure signal $f_0$ is the sum of the pseudo-Voigt shaped peaks \cite{Voight}, which have different  positions,  amplitudes and widths. Each peak $k$ can be mathematically described as follows
\begin{equation}
f_k(\nu_i)=\eta_k L_k(\nu_i)+(1-\eta_k)G_k(\nu_i)
\end{equation}
with $1\leq i\leq 800$, where
\begin{equation}
\begin{array}{ccc}
L_k(\nu_i)=\frac{\gamma_k}{\pi((\nu_i-c_k)^2+\gamma_k^2)},& & 
G_k(\nu_i)=\frac{1}{\sqrt{2\pi}\gamma_k}\exp\left(\frac{-(\nu_i-c_k)^2}{2\gamma_k^2}\right)
\end{array}
\end{equation}
 and 
\begin{equation} 
f_0(\nu_i)= \sum_{k=1}^M\alpha_kf_k(\nu_i)
\end{equation}
is the simulated pure spectrum, where $M$ is the total number of peaks. The parameters $c_k$ is the peak position, $\alpha_k$ is the maximum intensity, $\gamma_k$ and $\eta_k$ are the shape control parameters. The parameters used in this study are $c_1=800$, $c_2=850$, $\alpha_1=0.8$, $\alpha_2=0.2$, $\gamma_1=4.9$, $\gamma_2=7.1$, $\eta_1=0$ and $\eta_2=0.5$.

The baseline, modeled as a polynomial function, can be expressed as
\begin{equation}
{\bf b} = {\bf T}{\bf a}
\end{equation}
where ${\bf T}$ and ${\bf a}$ are defined as Vandermonde matrix and polynomial coefficients, respectively. The elements of Vandermonde matrix are given by $T_{i,j}=\nu_i^j$, with $4\geq j\geq 0$. The column vector ${\bf a}$ has length equal to $p+1$ for a degree of the polynomial function $p=4$. The polynomial coefficients used in this study are $a_1=1.0$, $a_2=0.027$, $a_3=0.16$, $a_4=0.2$ and $a_5=1.0$. 

Finally, random noise was also added to the simulated spectra. The experimental noise are modeled here as random numbers which are generated from Gaussian distribution function of mean $\mu$ and standard deviation $\sigma$. The noise function ${\bf n}$ can be mathematically described as follows
\begin{equation}
N(\mu, \sigma)\rightarrow n(\nu_i) 
\end{equation}
where $N$ denotes distribution function from which $n(\nu_i)$ is obtained. The parameter that allowed to manipulate the final level of noise was the standard deviation $\sigma$, in this study $\sigma=0.02$.   

\subsection{Denoising results}

\begin{center}
\begin{figure}[h]
\centering
\includegraphics[scale=.7]{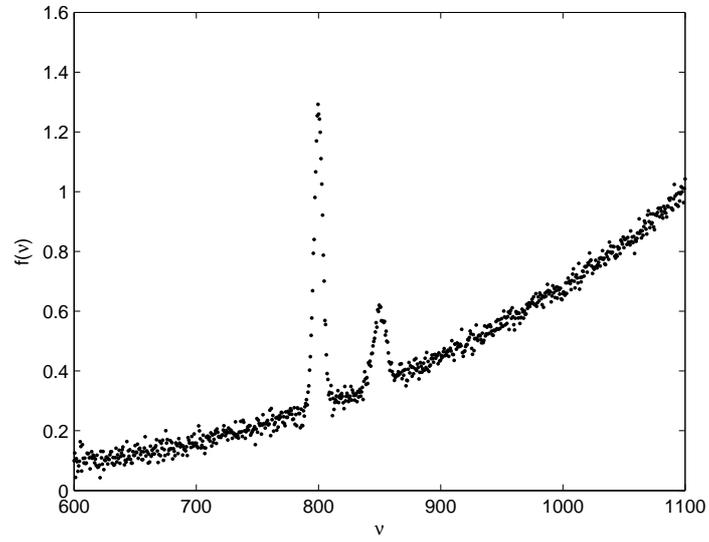}
\caption{The simulated spectra ${\hat{\bf f}}$, of range of $600$ to $1100$ cm$^{-1}$, with two peaks, baseline and noise.}
\label{fig1}
\end{figure}
\end{center}

The simulated spectra $\hat{{\bf f}}={\bf f}_{exp}$ with two peaks, baseline and noise, shown in Figure \ref{fig1}, was built according to the subsection \ref{simulate} in the range of $600$ to $1100$ cm$^{-1}$. The dimension of column vector $\hat{{\bf f}}$ is $800\times 1$. In the next step, the clear spectrum is determined by Equation (\ref{eqtik2}).

\begin{center}
\begin{figure}[h]
\centering
\includegraphics[scale=.7]{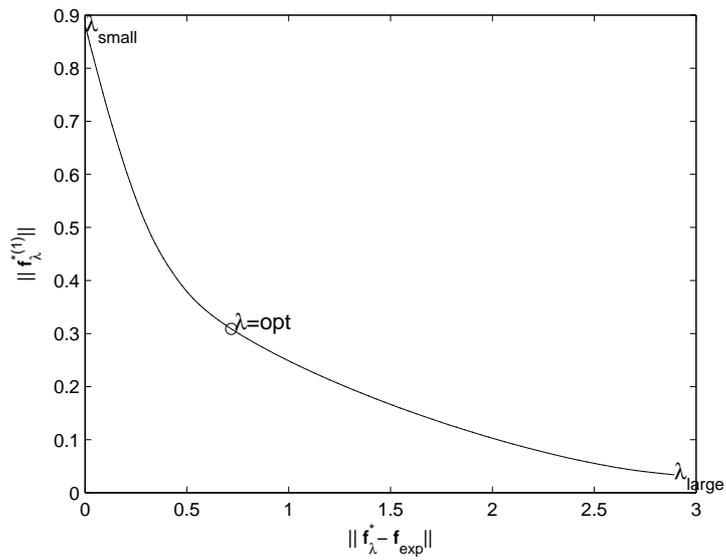}
\caption{The norm of the first-order derivative of the regularized function versus the norm of the 
corresponding difference between regularized and original function. Each point of the curve corresponds to different values of regularization parameter and the circle determines the point that corresponds to $\lambda=3.16$. All points of the curve were obtained with $\alpha=1$ in Equation (15).}
\label{fig2}
\end{figure}
\end{center}

If this problem is solved without Tikhonov regularization (Equation (\ref{eqtik2}), with $\lambda=0$), we find  the original spectrum as solution, i.e. $\bf{ f}^*_0=\hat{\bf{ f}}$, that has a large norm of the first-order derivative. If lower regularization parameter is chosen, there will be a small decrease  in the amplitude of the oscillations and the norm of the first-order derivative is smaller. This lowering  in oscillations increases with the value of the regularization parameter. On the other hand, if larger values of regularization parameter are considered, it is observed a distance from  shape of the original spectrum $\hat{{\bf f}}$, losing much details which could be correlated to some physical or chemical property of interest. Then we can control the shape of recovered function ${\bf f}_{\lambda}^*$ by choosing regularization parameter $\lambda$. The regularization parameter was chosen initially using modified L-curve method (Figure \ref{fig2}).

The modified L-curve is a plot of the norm of the first-order derivative of the regularized function, 
$||{\bf f}_{\lambda}^{*(1)}||$ versus the norm of the corresponding difference between regularized and original function, $||{\bf f}_{\lambda}^*-\hat{{\bf f}}||$. Each point on the curve corresponds to different values of regularization parameter. In L-curve, the corner point (maximum curvature) corresponds to a probable good choice for the regularization parameter, where there is a balance between the $||{\bf f}_{\lambda}^{*(1)}||$ and $||{\bf f}_{\lambda}^*-\hat{{\bf f}}||$. However, the value obtained by the maximum curvature criterion, which is an initial choice for $\lambda$ (in this case $\lambda=0.06$), may be modified based on Morozov's discrepancy principle. From discrepancy principle the best value of the regularization parameter  is such that $||{\bf f}_{\lambda}^*-\hat{{\bf f}}||$ is close to experimental error. After analyzing the obtained results, the chosen regularization parameter was equal to $\lambda=3.16$, in which $||{\bf f}_{\lambda}^*-\hat{{\bf f}}||$ corresponds to the maximum tolerable value. The point on the modified L-curve that corresponds to regularization parameter 3.16 is shown in Figure \ref{fig2} as a circle symbol, in which $||{\bf f}_{3.16}^{*(1)}||=3.10\times 10^{-1}$ and $||{\bf f}_{3.16}^*-\hat{{\bf f}}||=7.23\times 10^{-1}$. Figure \ref{fig2} also shows the values of $||{\bf f}_{\lambda}^{*(1)}||$ and $||{\bf f}_{\lambda}^*-\hat{{\bf f}}||$ for small and large regularization parameter. The ideal function ${\bf f}_{\lambda}^*$ provides a small value to $||{\bf f}_{\lambda}^{*(1)}||$ while providing a small value to $||{\bf f}_{\lambda}^*-\hat{{\bf f}}||$.

\begin{center}
\begin{figure}[h]
\centering
\includegraphics[scale=.7]{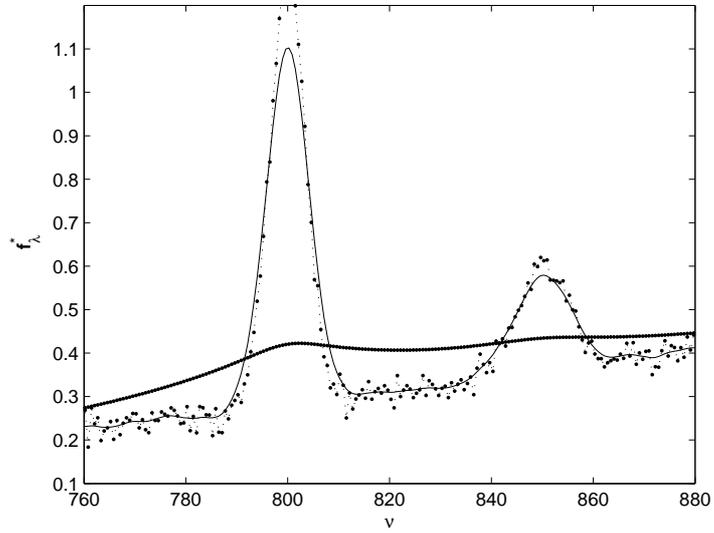}
\caption{Recovered function to different values of regularization parameter, of range of $10^{-4}$ to $10^3$. Recovered function with $\lambda=3.16$ is showed as continuous line, $\lambda=10^{-4}$ as dash-dotted line and $\lambda=10^3$ as dotted line.  All results were obtained with $\alpha=1$ in Equation (15).}
\label{fig8}
\end{figure}
\end{center}

Using Equation (\ref{eqtik2}) to find ${\bf f}_{\lambda}^*$, with $\lambda=3.16$ ($a_{1}=1$ and $a_{2}=0$), we arrived at a result with few oscillations. This result is shown in Figure \ref{fig8} (continuous line). Figure \ref{fig8} also shows the recovered function with a smaller value ($\lambda=1\times 10^{-4}$, dash-dotted line) and  larger value ($\lambda=1\times 10^{3}$, dotted line) of the regularization parameter. As observed in Figure \ref{fig2}, with smaller value of the $\lambda$ parameter, the norm of the first-order derivative of the function is large (i.e. with great amplitude in the oscillations). On the other hand, with greater value of the $\lambda$ parameter, the difference between original function and recovered function is big. In this case, even for small amplitude in oscillations, the recovered function is not suitable. These results can be confirmed in Figure \ref{fig8}. All results shown in Figure \ref{fig8} can be obtained with $\alpha=1$ in Equation (\ref{eqprincipal}). 

This paper proposes a different way to obtain recovered function without noise. When the Equation (\ref{eqprincipal})  is used to find ${\bf f}_{\lambda,\alpha}^*$, we have a new parameter, that is $\alpha$. In this case, with fixed $\lambda$ ($\lambda=3.16$), we can use $\alpha$ to control the amplitude of the oscillations. For the usual Tikhonov method, the value of $\alpha$ is equal to one. Therefore, there are many solutions that can be found using the Equation (15) but that cannot be found using the Equation (12).

\begin{center}
\begin{figure}[h]
\centering
\includegraphics[scale=.7]{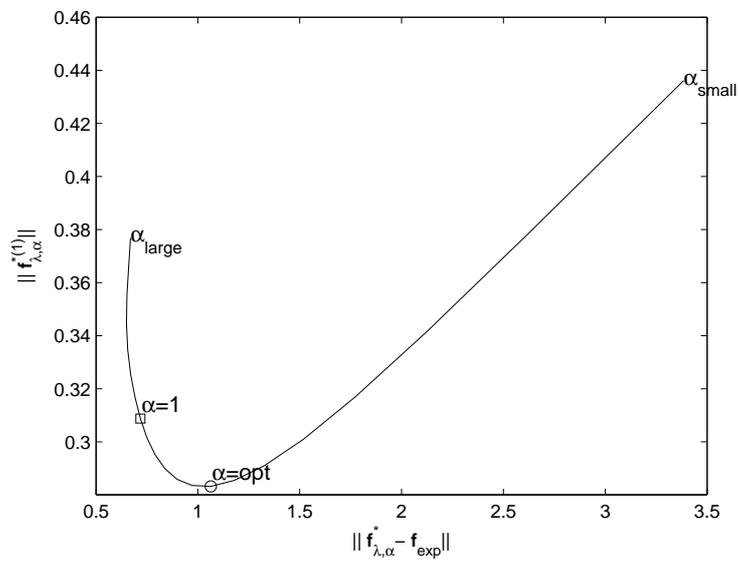}
\caption{The norm of the first-order derivative of the regularized function, 
versus the norm of the corresponding difference between regularized and original function. Each point of the curve corresponds to different values of fractional derivative order, with regularization 
parameter fixed ($\lambda=3.16$) in Equation (15).}
\label{fig4}
\end{figure}
\end{center}

The choice of $\alpha$ was made by using the graphical method (modified L-curve). The proposed graph is a plot of the norm of the first-order derivative of the regularized function, $||{\bf f}_{\lambda={\rm constant},\alpha}^{*(1)}||$ versus the norm of the corresponding difference between regularized and original function, $||{\bf f}_{\lambda={\rm constant},\alpha}^*-\hat{{\bf f}}||$. In this case, each point corresponds to different values of fractional derivative order, with regularization parameter fixed. When $\alpha$ is equal to zero, the recovered function ${\bf f}_{\lambda=3.16,0}^{*}$ does not have a small value of the difference norm (see small values of the $\alpha$ mark in Figure \ref{fig4}). In addition, the norm of first-order derivative is large too,  indicating the presence  of large oscillations (see dotted line in Figure \ref{fig7}). On the other hand, when $\alpha$ is large, we find a solution which comes close to the simulated value (see large values of the $\alpha$ mark in Figure \ref{fig4}), with a big oscillation (see dash-dotted line in Figure \ref{fig7}). Therefore, there is one $\alpha$ that establishes the good balance between the difference norm $||{\bf f}_{\lambda,\alpha}^*-\hat{{\bf f}}||$ and oscillations $||{\bf f}_{\lambda,\alpha}^{*(1)}||$, this optimum $\alpha$ provides the desired clear spectrum. The corner of the modified L-curve (Figure 4) provides the optimum value of fractional derivative order. 

\begin{center}
\begin{figure}[h]
\centering
\includegraphics[scale=.7]{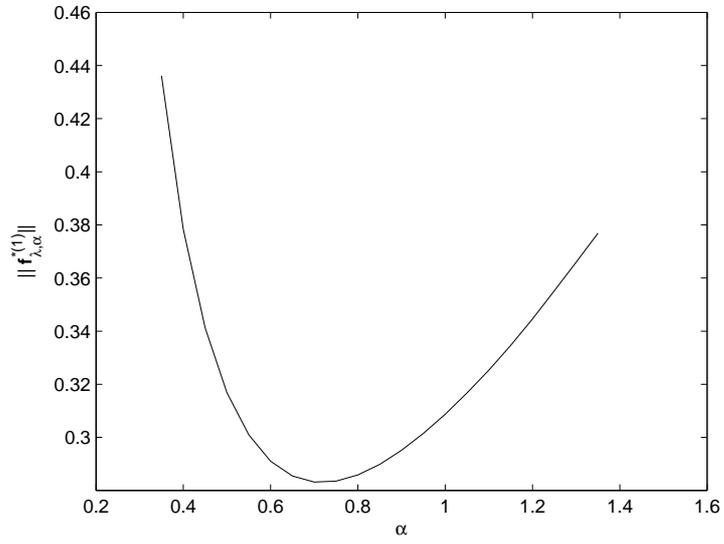}
\caption{The norm of the first-order derivative of the regularized function (Equation (15)) versus fraction order.}
\label{fig7xx}
\end{figure}
\end{center}

\begin{center}
\begin{figure}[h]
\centering
\includegraphics[scale=.7]{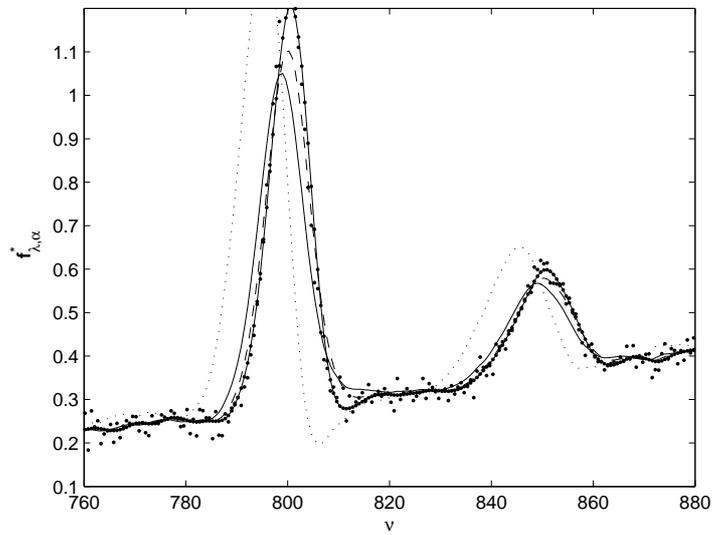}
\caption{Recovered function to different values  of derivative order, of range of $0.35$ to $1.35$. Recovered function with $\alpha=0.7$ is showed as continuous line, $\alpha=0.35$ as dotted line and $\alpha=1.35$ as dash-dotted line. All results were obtained with regularization parameter fixed ($\lambda=3.16$) in Equation (15). The dashed line shows the result with $\alpha=1$.}
\label{fig7}
\end{figure}
\end{center}

In order to clarify this point, the norm of the first-order derivative of the solution, $||{\bf f}_{\lambda=3.16,\alpha}^{*(1)}||$, is plotted in Figure \ref{fig7xx} for different values of $\alpha$. 
We suggest that the minimum of the curve in Figure \ref{fig7xx} provides an optimum value of $\alpha$, $\alpha^*$. This value is close to the value found in modified L-curve (Figure 4).

With $\alpha=\alpha^*=0.7$, we find a solution that has a small residual norm as well as oscillations with small amplitude, with $||{\bf f}_{3.16,0.7}^{*(1)}||=8.03 \times 10^{-2}$ and $||{\bf f}_{3.16,0.7}^*-\hat{{\bf f}}||=1.13$, respectively. When using $\alpha=1$, the results found were 
$||{\bf f}_{3.16,1}^{*(1)}||=3.10\times 10^{-1}$ and $||{\bf f}_{3.16,1}^*-\hat{{\bf f}}||=7.23\times 10^{-1}$. These results are shown in Figure \ref{fig4} with circle symbol for $\alpha=0.7$ and square symbol for $\alpha=1$. Therefore, the use of Equation (\ref{eqprincipal}) with $\alpha=0.7$ instead of $\alpha=1$ led us to a better result, when compared to the norm of first-order derivative of the solution, $||{\bf f}_{3.16,0.7}^{*(1)}||=8.03 \times 10^{-2}$ and $||{\bf f}_{3.16,1}^{*(1)}||=3.10\times 10^{-1}$. However the result found with $\alpha=0.7$ deviates further from the original value when compared with $\alpha=1$, $||{\bf f}_{3.16,0.7}^*-\hat{{\bf f}}||=1.13$ and $||{\bf f}_{3.16,1}^*-\hat{{\bf f}}||=7.23\times 10^{-1}$. 

Figure \ref{fig7} shows the recovered functions with a smaller value ($\alpha=0.35$, dotted line) and  larger value of the derivative order ($\alpha=1.35$, dash-dotted line). All results shown in Figure \ref{fig7} were obtained with $\lambda=3.16$ in Equation (\ref{eqprincipal}). The recovered functions with $\alpha=0.7$ ({continuous} line) and $\alpha=1$ ({dashed} line), both with $\lambda=3.16$, appear to have good similarity to the free noise spectrum, which does not compromise the final analysis. However, the recovered function with $\alpha=0.7$ oscillate less than with $\alpha=1$. 
This figure shows how  the fractional order $\alpha$ can help controlling the oscillation in 
recovered function.
 
\begin{center}
\begin{figure}[h]
\centering
\includegraphics[scale=.7]{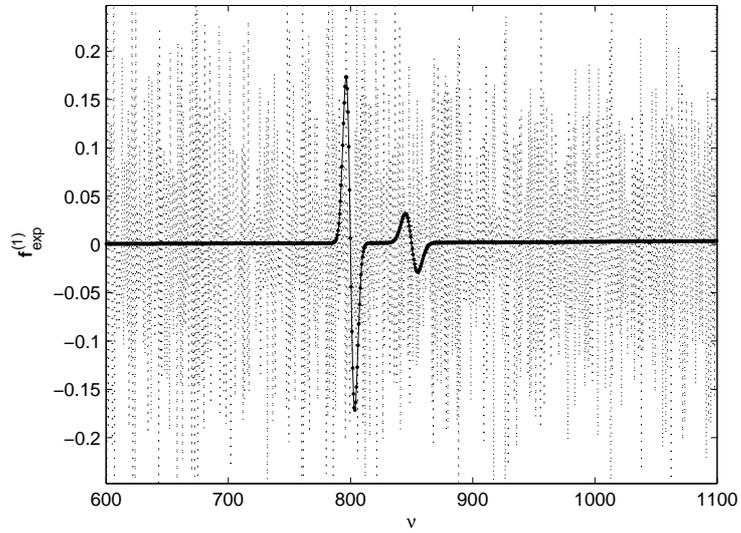}
\caption{First-order derivative of the function $\hat{{\bf f}}$ using operator ${\bf D}$, presented in Equation (11), without regularization (dotted line). The dash-dotted  line shows the expected result.}
\label{fig8b}
\end{figure}
\end{center}

\begin{center}
\begin{figure}[h]
\centering
\includegraphics[scale=.7]{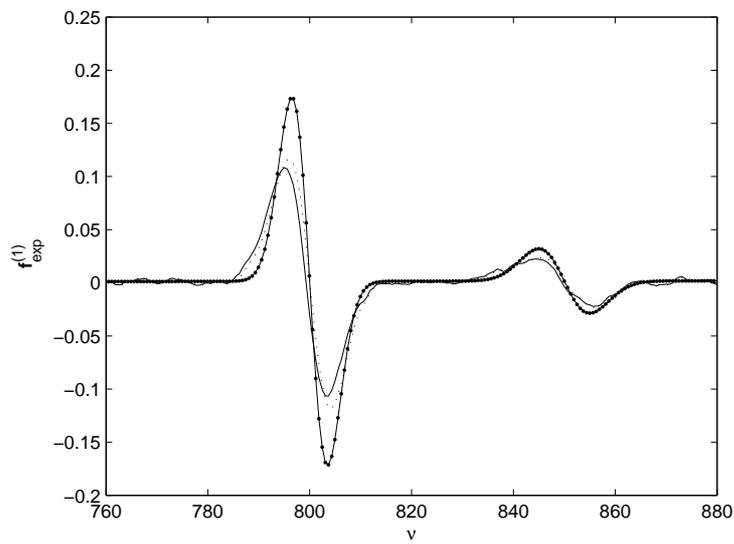}
\caption{First-order derivative of the function $\hat{{\bf f}}$ with regularization, $\alpha=0.8$ and $\lambda=3.16$ (continuous line). The dash-dotted line shows the expected result and dotted line shows the function with $\alpha=1$ and $\lambda=3.16$. All results were obtained using Equation (17).}
\label{fig8c}
\end{figure}
\end{center}

The determination of the first-order derivative of the function $\hat{{\bf f}}$ is a difficult problem to solve, when  experimental errors are taken into account.
The difficulties in finding the first-order derivative of the function $\hat{{\bf f}}$ using operator ${\bf D}$ (Equation (11)) are shown in Figure \ref{fig8b}. As we can see, the usual solution procedure (Equation (11)) is unstable, which leads to large errors in the final results. Due this fact, this problem is called ill-posed \cite{Tikhonov1}. The usual way to circumvent this difficulty is presented in Equation (16), in which Tikhonov strategy is used. The Equation (17) was proposed here, in order to introduce the fractional derivative tools in calculation of the first-order derivative, in the same way we did in the filter proposal. Figure \ref{fig8c} shows the result obtained by applying Equation (17) with $\lambda=3.16$ and $\alpha=0.8$. The use of the Equation (17) provides a stable solution and free of oscillations that does not introduce false peaks into the result. Differentiation procedure reveals details of slope change in the experimental spectrum that can reveal  the presence of hidden and overlapped peaks.

\begin{center}
\begin{figure}[h]
\centering
\includegraphics[scale=.7]{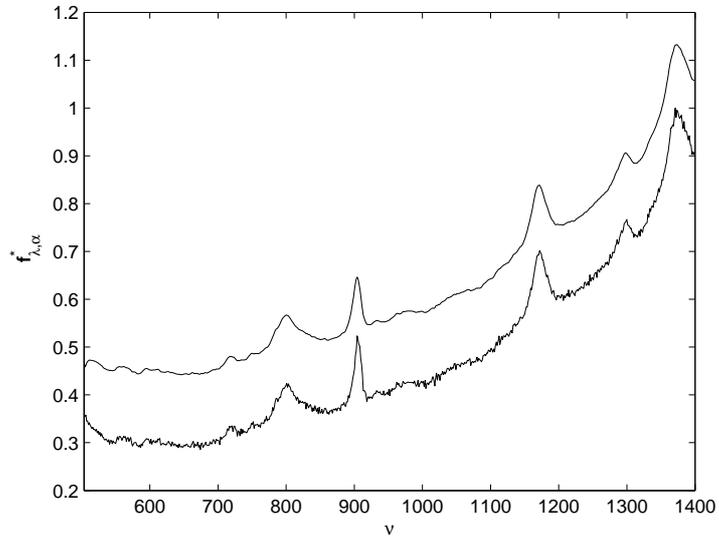}
\caption{Experimental data spectra (bottom line) and clear spectra (top line) obtained by Equation (15) with $\alpha=5.01$ and $\lambda=0.6$. The clear spectra was shifted in the vertical direction 
to facilitate comparison with experimental data.}
\label{fig10}
\end{figure}
\end{center}

In our next example, we will analyze a real spectra. Now the proposed method was applied to surface-enhanced Raman scattering (SERS) spectra of crystal violet dye in Ag nanoparticles colloidal dispersion. The details of experimental procedure and synthetic steps are specified in section 3. The experimental spectra, shown in Figure \ref{fig10}, was obtained 
in the range of $503$ to $1400$ cm$^{-1}$. In this case, the dimension of column vector $\hat{{\bf f}}$ is $1388\times 1$. Initially, experimental spectra was normalized, such that $\hat{\bf f}=\hat{\bf f}/max(\hat{\bf f})$. Figure \ref{fig10} shows experimental spectra and clear spectra obtained using Equation (15) with $\lambda=5.01$ and $\alpha=0.6$. The values of $\lambda$ and $\alpha$ were obtained following the methodology described before. Again, the result obtained with fractional $\alpha$ derivative order is most satisfying, presenting small oscillations ($||{\bf f}_{5.01,0.6}^{*(1)}||=0.079$) when compared to $\alpha=1$ ($||{\bf f}_{5.01,1}^{*(1)}||=0.086$).

Based on these examples, we can say that the proposed method offers a promising approach to remove noise effects in experimental data and first-order derivative of the experimental data.

\section{Conclusion}

Raman spectroscopy is a well-established technique that allows chemical  analysis of materials. The existence of random noise in experimental data can negatively affect this analysis. As seen in our discussion, the first derivative of the experimental data is very sensitive to the noise. Therefore, the random noise should be fitted to minimize chance of misinterpretation. In this paper, the random noise of the Raman signal was suppressed by modified Tikhonov method, in which fractional-order derivative of the solution was used as additional criterion in Tikhonov function. 

This strategy shows more versatility due to new parameter added. Out of special cases in which $\alpha=0,1$ or $2$, the fractional Tikhonv method provides many solutions which are not found by usual Tikhonov procedure. To investigate performance of the proposed method, we compared it with original Tikhonov method. In the first example, simulated data was considered to study the proposed method. In this case, it was found better results using $\alpha=0.7$ instead of $\alpha=0,1$ or $2$.
This result was achieved even with the presence of experimental errors (coefficient of variation of $7.4\%$) in simulated data. The proposed method was applied successfully to two other examples: the determination of the first-order derivative of the experimental data and SERS spectra of crystal violet dye in Ag nanoparticles colloidal dispersion. Once more, in both cases, better results were found with fractional order different than $\alpha=1$.

The choice of $\alpha$ was made by using the graphical method, in which Euclidian norm of the first-order derivative of the solution  is plotted in function of the norm of the difference between the analytical signal and recovered signal, for different values of $\alpha$. By this way, the minimum of the curve  provides a optimum value of $\alpha$. Future works should explore different strategies to found optimal $\alpha$ parameter.

Although there is no significant difference between $\alpha=1$ and fractional derivative order, because, in the examples presented earlier, usual Tikhonov is a suitable method to achieve excellent results, the FTM method proved to be an interesting tool that should be better explored in other cases.

\section*{Acknowledgments}

This study was financed in part by the Coordena\c c\~ao de Aperfei\c coamento de Pessoal de N\'ivel Superior - Brasil (CAPES) - Finance Code 001 and FAPEMIG. 

\clearpage

\clearpage

\clearpage

\clearpage

\clearpage

\clearpage

\bibliographystyle{plain}

\end{document}